\documentclass[prl,two column]{revtex4}
\usepackage[dvips]{graphicx}
\usepackage{subfigure}
\usepackage{bm}
\usepackage{color}
\usepackage{amsmath,amssymb}
\begin{document}
\title{ Spinmotive force due to motion of magnetic bubble arrays\\ driven by magnetic field gradient}
\author{Y. Yamane$^1$, S. Hemmatiyan$^2$, J. Ieda$^3$, S. Maekawa$^3$, and J. Sinova$^{1,2,4}$}
\affiliation{1 Institut f\"{u}r Physik, Johannes Gutenberg Universit\"{a}t Mainz, 55128 Mainz, Germany}
\affiliation{2 Department of Physics, Texas A\&M University, College Station, Texas 77843-4242, USA}
\affiliation{3 Advanced Science Research Center, Japan Atomic Energy Agency, Tokai,
Ibaraki 319-1195, Japan}
\affiliation{4 Spin Phenomena Interdisciplinary Center (SPICE), Johannes Gutenberg Universit\"{a}t Mainz, 55128 Mainz, Germany}
\maketitle
{\bf
Interaction between local magnetization and conduction electrons is responsible for a variety of phenomena in magnetic materials.
It has been recently shown that spin current and associated electric voltage can be induced by magnetization that depends on both time and space.
This effect, called spinmotive force, provides for a powerful tool for exploring the dynamics and the nature of magnetic textures, as well as a new source for electromotive force.
Here we theoretically demonstrate the generation of electric voltages in magnetic bubble array systems subjected to a magnetic field gradient.
It is shown by deriving expressions for the electric voltages that the present system offers a direct measure of phenomenological parameter $\beta$ that describes non-adiabaticity in the current induced magnetization dynamics.
This spinmotive force opens a door for new types of spintronic devices that exploit the field-gradient.
}

Spinmotive force (SMF) refers to the generation of spin current, which is accompanied by an electric voltage, as a result of dynamical magnetic textures in conducting ferromagnets\cite{stern,barnes,review}.
This is due to the exchange coupling between conduction electrons and the local magnetization.
SMF reflects the temporal- and spatial-dependence of the local magnetization\cite{volovik,duine,tserkov,lee,tatara,prb}, and thus it offers a powerful method to probe and explore the dynamics and the nature of magnetic textures.
In addition, SMF can be a new source of electromotive force, directly converting the magnetic energy into the electric energy of conduction electrons.
While the classical electromagnetism tells us that the conventional inductive electromotive force requires a time-varying magnetic flux, it has been reported that an electromotive force can be generated by a static and uniform magnetic field via the SMF mechanism\cite{yang,hai,hayashi}.

\begin{figure}[b]
    \centerline{\includegraphics[width=80mm]{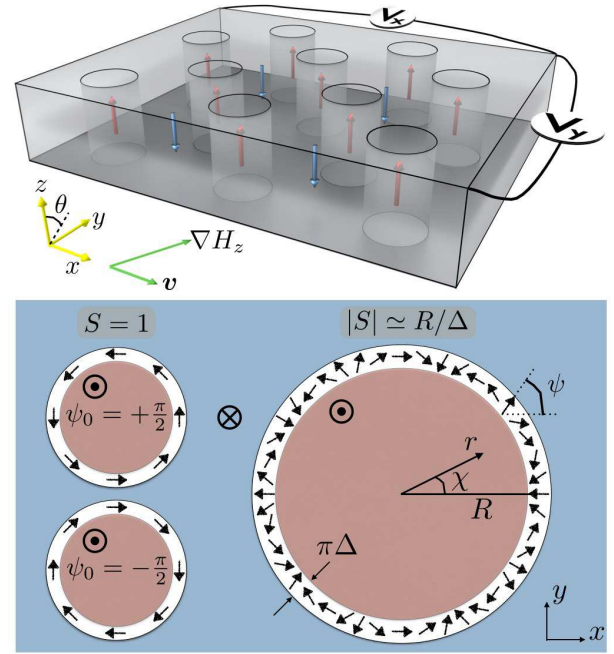}}
    \caption{  Schematic of magnetic bubble structures in a thin film.
Red and Blue arrows in the upper figure and black ones in the bottom indicate the magnetization.
            }
    \label{fig3}
\end{figure}

In the last few years, more attention has been focusing on topologically nontrivial magnetic structures such as magnetic vortices in soft ferromagnetic nanodiscs\cite{vortex1,vortex2} and skyrmion lattices in chiral magnetic thin films\cite{skyrmion1,skyrmion2}.
The SMF offers some insights into and gains benefit from such magnetic systems;
the polarity of a moving magnetic vortex core can be electrically detected\cite{ohe2009,tanabe}, and arbitrarily-large ac SMF was predicted due to skyrmion lattice motion\cite{ohe2013}.
To the best of our knowledge, however, there has been no work on SMF induced in systems that contain magnetic bubble domains.
Magnetic bubbles are observed in ferromagnetic films with out-of-plane anisotropy as spot-like closed domains, where the magnetization is oriented to the opposite direction to the one outside the bubbles.
The structures of magnetic bubbles are similar to those of skyrmions and vortices in the sense that bubbles carry a topological number called skyrmion number.
Since their first observation in the 1960s, magnetic bubbles have been showing distinctive and interesting behaviors\cite{text,bubble1,bubble2,bubble3}.

In this paper, SMF due to the motion of magnetic bubble arrays is theoretically investigated based on the steady-motion model.
As the bubble motion may be induced by a spatially-varying magnetic field, our work reveals that a magnetic field {\it gradient} can be exploited to generate a spin current and an associated electric voltage.
By deriving expressions for the electric voltages, we demonstrate their cumulative nature, i.e., they can be proportional to the number of involved bubbles.
An important fact to be stressed is that the present system can confirm the SMF originating from the non-adiabatic dynamics of the electron spin, leading to a direct measurement of the controversial, so-called, $\beta$ parameter.
\\

\noindent {\bf Steady-motion model for magnetic bubble.}
Let us begin by reviewing the collective-coordinate model for magnetic bubble dynamics.
We consider a cylindrical bubble domain with radius $R$ in a thin film.
This type of magnetic structure can be stabilized by applying bias magnetic field in the out-of-plane with appropriate magnitude.
The distribution of the magnetization direction ${\bm m}=(\sin\theta\cos\psi, \sin\theta\sin\psi,\cos\theta)$ is assumed to be two dimensional and described as\cite{text} [see Fig.~1]
\begin{eqnarray}
\theta\left(r,\chi,z\right) &=& 2\tan^{-1} \left[ \exp\left(\frac{Q(r-R)}{\Delta}\right) \right] \label{theta}\\
\psi\left(r,\chi,z\right) &=& S\chi + \psi_0\label{psi}
\end{eqnarray}
where $(r,\chi,z)$ is the cylindrical coordinate measured from the bubble center, $\Delta$ is the wall width parameter, $\psi_0$ is a constant, $Q$ is the topological parameter defined as
\begin{equation}
Q=\frac{1}{\pi}\int_0^\infty \frac{\partial\theta}{\partial r}dr=\pm1
\end{equation}
and $S$ is the winding number:
\begin{equation}
S=\frac{1}{2\pi}\int_{\chi=0}^{2\pi}d\psi = \frac{1}{2\pi}\oint\frac{d\psi}{ds}ds,
\end{equation}
where $\oint ds$ is the contour integral taken counterclockwise around the circumference.
The domain wall separating the bubble and the outside, in general, can contain vertical Bloch lines, i.e., there are many possibilities in the way of distributing the azimuthal angle $\psi$ along the perimeter.
In Eq.~(\ref{psi}) we assumed the linear dependence of $\psi$ on $\chi$,
as we focus on the following two cases.
First, when the magnetization rotates one full turn around the wall of the bubble with no Bloch line, $\psi=\chi\pm\pi/2$, that is, $S=1$ and $\psi_0=\pm\pi/2$, with the sign $+$ ($-$) corresponding to left (right) handed chirality.
Eq.~(\ref{psi}) is also a good approximation when the Bloch lines are packed so closely that the distance between the adjacent
Bloch lines is comparable to the wall width, i.e., $|S|\simeq R/\Delta$.
If one considers a small number of Bloch lines, the distribution of $\psi$ would be no longer as simple as Eq.~(\ref{psi}).

When a magnetic field is applied in the $z$ direction with its magnitude varying in the $x$-$y$ plane, a bubble is driven to move in the film seeking positions with lower Zeeman energy.
In the following the steady motion of the bubble under the constant gradient $\nabla H_z$ is assumed, i.e., during its motion with constant velocity ${\bm v}$ the bubble stays rigidly cylindrical with constant radius $R$ and the $\psi$-distribution does not change with respect to
the coordinate frame moving with the bubble:
\begin{equation}
\theta({\bm x},t) = \theta({\bm x}-{\bm v}t), \qquad \psi({\bm x},t) = \psi({\bm x}-{\bm v}t).
\label{steady}\end{equation}
Assuming Eq.~(\ref{steady}) and that the magnetization dynamics obeys Landau-Lifshitz-Gilbert equation, the equation of motion for the bubble is given by\cite{text} (the derivation is shown in Appendices)
\begin{equation}
\nabla H_z
= \frac{2S}{R^2\gamma} \left[\frac{Q\alpha}{2}\frac{R}{S\Delta}\left( 1 + \frac{S^2\Delta^2}{R^2}\right){\bm v}+{\hat{\bm z}}\times{\bm v}\right],
\label{eom}\end{equation}
where $\gamma$ is the gyromagnetic ratio and $\alpha$ is the dimensionless Gilbert damping parameter.
$\hat{\bm{z}}$ represents the unit vector along the $z$-direction.
An assumption that was made when deriving Eq.~(\ref{eom}) is that the spin transfer with the conduction electrons are negligible.
Remarkably, the bubble is deflected away from the direction of the field gradient at an angle that is determined by the material parameters (see Appendices for the detail).
The extension of the above discussion to multiple-bubble problem is straightforward.
\\

\noindent {\bf Spinmotive force due to bubble motion.}
Let us examine the SMF induced by the steady motion of bubble indicated by Eq.~(\ref{steady}).
We assume that a conduction electron in the ferromagnetic film is described by a one-body Hamiltonian
\begin{equation}
\mathcal{H} = \frac{\bm{p}^2}{2m_{\rm e}} + J_\mathrm{ex} \bm{\sigma}\cdot\bm{m}({\bm r},t),
\label{h}\end{equation}
where $m_e$ is the electron's mass.
The second term represents the exchange interaction between the electron spin and the magnetization, with $J_{\rm ex}$ being the exchange coupling energy.
According to theory of SMF\cite{volovik,duine,tserkov}, dynamical magnetization exerts an effective electric field $\pm\bm{{\cal E}}$ on the electrons via the exchange coupling, which is called spin electric field since its sign depends on the electron spin (see Appendices):
\begin{equation}
\pm\bm{{\cal E}} = \pm\left( \bm{{\cal E}}^{\rm A} + \bm{{\cal E}}^{\rm NA}\right),\label{es}
\end{equation}
with
\begin{eqnarray}
\bm{{\cal E}}^{\rm A} &=& \frac{\hbar}{2e}\sin\theta\left( \frac{\partial\theta}{\partial t}\nabla\psi - \frac{\partial\psi}{\partial t}\nabla\theta \right),\label{es-ad}\\
\bm{{\cal E}}^{\rm NA} &=&  \beta\frac{\hbar}{2e}\left( \frac{\partial\theta}{\partial t}\nabla\theta + \sin^2\theta \frac{\partial\psi}{\partial t}\nabla\psi \right).\label{es-nad}
\end{eqnarray}
The upper (lower) signs in Eq.~(\ref{es}) correspond to the electron with majority (minority) spin.
$\bm{{\cal E}}^{\rm A}$ and $\bm{{\cal E}}^{\rm NA}$ are referred to as adiabatic and non-adiabatic spin electric fields, respectively, as $\beta=\hbar/2J_{\rm ex}\tau_{\rm sf}$ is the dimensionless parameter describing the non-adiabaticity in the electron spin dynamics\cite{duine,tserkov,tatara,prb}, with $\tau_{\rm sf}$ the relaxation time for the electron spin flip.
The spin electric fields~(\ref{es-ad}) and (\ref{es-nad}) require both time and spatial dependences of the magnetization, and this condition is satisfied around the perimeter of moving bubbles.

The spin electric field (\ref{es}) induces a spin current ${\bm j}_{\rm s}=-(\sigma_F^\uparrow + \sigma_F^\downarrow)\bm{{\cal E}} $ and a charge current ${\bm j}_c = (\sigma_F^\uparrow - \sigma_F^\downarrow)\bm{{\cal E}}$ in the sample, with $\sigma_F^{\uparrow(\downarrow)}$ the electric conductivity for the majority (minority) electrons.
These currents generate by-products such as the charge redistribution, the spin accumulation, and the charge/spin diffusion current.
In an open circuit system, the electric field ${\bm E}_{\rm ind}=-\nabla\phi-\partial{\bm A}/\partial t$ appears to keep the total charge current zero, i.e., ${\bm j}_c + (\sigma_F^\uparrow+\sigma_F^\downarrow){\bm E}_{\rm ind}=0$, where $\phi$ and ${\bm A}$ are the electromagnetic scalar and vector potentials.
An electric voltage between two given points ${\bm x}_a$ and ${\bm x}_b$, which is given by the difference in the electric potential $\phi({\bm x}_b) - \phi({\bm x}_a)$, enables one to detect the spinmotive force electrically. 
To determine the gauge potentials one has to fix the gauge, and here let us adopt the Coulomb gauge, $\nabla\cdot{\bm A}=0$.
From the above equation for the open circuit condition, one obtains the Poisson equation
\begin{equation} 
-\nabla^2\phi=\nabla\cdot{\bm F}.
\label{poisson}\end{equation}
Here ${\bm F}=-\nabla\phi=-P\bm{{\cal E}}$ is the conservative electric field induced by the electric potential distribution, where $P=(\sigma_F^\uparrow - \sigma_F^\downarrow)/(\sigma_F^\uparrow + \sigma_F^\downarrow)$ is the spin polarization of the conduction electrons.
In the above argument, we have neglected the contribution from the diffusive current to the total charge current as we have metals in our mind as the samples;
once the effective U(1) electric field $\bm{{\cal E}}$ is given, the problem of computing the electric voltage induced by the electric field falls within the classical electromagnetism and the established transport theory, and it is known that in metals the induced diffusion potential is mostly negligible compared to the electric potential, unlike in semiconductors.
Technically, the diffusive current can be taken into account by replacing the electric potential $-e\phi$ by the electrochemical potential $\mu=-e\phi + \epsilon_F$, where $\epsilon_F$ is the Fermi energy.
($\epsilon_F$ may be dependent on the space and the spin electric field in complex ways.)

Eq.~(\ref{poisson}) can be applied to systems with arbitrary sample geometry and magnetic texture.
In Fig.~2, we show an example of electric potential distribution by numerically solving the Poisson equation~(\ref{poisson}) with spin electric fileds (\ref{es-ad}) and (\ref{es-nad}), where the steady motion of eight identical bubbles in a square-shape thin film is assumed [see Appendices for the numerics].
Here the coordinate system is set that the bubble array flows along the $x$ direction.
It is seen that the potential drop occurs at the position of  the bubbles, as is expected.
Notice that the adiabatic field gives rise to the net potential drops only in the $y$ direction (perpendicular to the bubble flow), while the non-adiabatic one only to the $x$ direction (along the bubble flow), indicating that in this setup the two contributions can be separately identified by longitudinal and perpendicular voltage measurements. 
While $S=1$ is assumed in Fig.~2, qualitatively much the same profiles are also obtained in the case of $|S|=R/\Delta$, but with height of each potential drop being different (not shown).
The $S$ dependence of the electric voltage will be discussed later.

\begin{figure}[t]
    \centerline{\includegraphics[width=48mm]{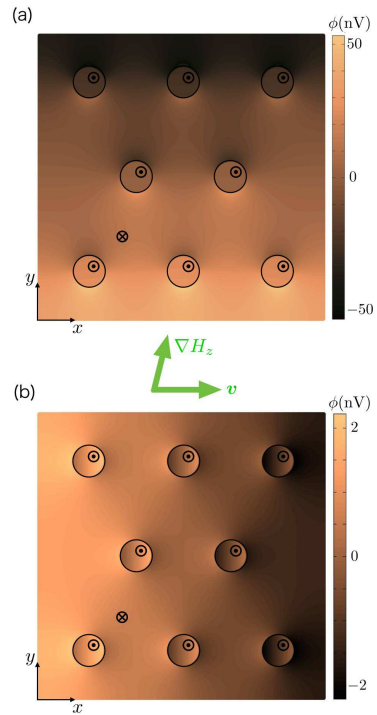}}
    \caption{  The distributions of electric potential $\phi$ induced in a thin film due to (a) the adiabatic field (\ref{es-ad}) and (b) the non-adiabatic field (\ref{es-nad}), calculated by solving the Poisson equation (\ref{poisson}) numerically, where the steady motion of eight identical bubbles along the $x$ direction is assumed.
    The profile of each bubble is given by Eqs.~(\ref{theta}) and (\ref{psi}), with $R=50$ nm, $\Delta=2$ nm, $Q=1$, $S=1$ and $\psi_0=\pi/2$.
    The other parameters assumed here are $\gamma=1.76\times10^{11}$ T$^{-1}$s$^{-1}$, $\alpha=0.02$, $\beta\simeq0.0033$, $P=0.5$, and the side lengths are $900$ nm.
    The field gradient makes an angle $75.9^\circ$ to the $x$ axis with its strength $|R\nabla H_z|=10$ Oe.
            }
    \label{fig2}
\end{figure}

Next, let us adopt a ``quasi-one-dimensional'' approximation for the electric voltage and derive its analytic expression.
We limit ourselves to a rectangular thin film where bubbles move along either of two sides of sample, as is the case in Fig.~2.
Assuming a bubble moving in the positive $x$ direction, we estimate the electric voltage $V_x$ along the motion of the bubble by
\begin{eqnarray}
V_x &\simeq& \frac{1}{L_y}\int_0^{L_y}dy\int_0^{L_x} \frac{\partial\phi}{\partial x} dx
= \frac{1}{L_y}\int\int rdrd\chi  P{\cal E}_x    \nonumber\\
&\simeq& -\frac{R}{L_y} \frac{P\hbar}{2e} \frac{1+\left(|S|\Delta/R\right)^2}{|S|\Delta/R}  \nonumber\\
&&\times
\left[1+\alpha^2\left(\frac{1+\left(S\Delta/R\right)^2}{2S\Delta/R}\right)^2\right]^{-1/2}
\pi\gamma\beta |R\nabla H_z|,\nonumber\\
\label{vx}\end{eqnarray}
where $L_{x(y)}$ is the side length of the sample along the $x(y)$ direction.
The $y$-integral operation and the appearance of the factor $1/L_y$ are for spatial-averaging of the electric potential along the $y$-axis.
Eqs.~(\ref{theta}), (\ref{psi}), (\ref{steady}), (\ref{eom}) and (\ref{es}) have been used to carry out the integrals.
Eq.~(\ref{vx}) gives the exact solution for $V_x$ in one-dimensional systems: $V_x=\int (\partial\phi/\partial x)dx$.
In two-dimensional systems as is the present case, Eq.~(\ref{vx}) is still a good approximation when effects on $V_x$ from $\partial\phi/\partial y$ and the detailed profile of $\partial\phi/\partial x$ can be ignored.
This is often the case if there are some appropriate symmetries in the system, and the divergence of the electric field, i.e., the moving bubble, keeps away from the electrodes.
Similar concept for the electric voltage $V_y$ measured in the perpendicular direction to the motion of the bubble leads to
\begin{eqnarray}
V_y &\simeq&  \frac{1}{L_x}\int_0^{L_x}dx\int_0^{L_y} \frac{\partial\phi}{\partial y} dy \nonumber\\
&\simeq&  -\frac{R}{L_x}\frac{P\hbar}{2e} \frac{S}{|S|}\nonumber\\
&&\times\left[1+\alpha^2\left(\frac{1+\left(S\Delta/R\right)^2}{2S\Delta/R}\right)^2\right]^{-1/2}
2\pi \gamma Q |R\nabla H_z|.\nonumber\\
\label{vy}\end{eqnarray}
The dc electric voltages appear both in the $x$ and $y$ directions, being proportional to the field gradient for both magnetic configurations $S=1$ and $|S|\simeq R/\Delta$.
It is seen from Eq.~(\ref{vx}) that $V_x$ for $S=1$ is larger than that for $|S|=R/\Delta$ under the same applied field, because $R>\Delta$.
With the parameters shown in Fig.~2, $V_x$ for $S=1$ is about an order of magnitude greater compared to that for $|S|=R/\Delta$.
On the other hand, $V_y$ has little dependence on $|S|$.

When there are multiple bubbles, the net electric potential distribution is given by the superposition of all individual electric potentials produced by each bubble.
If $N$ identical bubbles move in the $x$ direction, the simplest extensions, $V^N
_x$ and $V^N_y$, of the above expressions for the electric voltages, $V_x$ and $V_y$, respectively, may be
\begin{equation}
V_x^N= N V_x, \qquad V_y^N = N V_y.
\label{vn}\end{equation}
Figure~3 compares Eqs.~(\ref{vn}) with the electric voltages obtained by numerically solving the Poisson equation (\ref{poisson}), showing good agreement between them.
\\

\noindent {\bf Discussion}
Eqs.~(\ref{vx}), (\ref{vy}) and (\ref{vn}) indicate key features of this SMF.
$V_x^N$ is proportional to $N/L_y$, and thus, roughly speaking, depends on the ``density'' of bubbles along the $y$ axis and the ``number'' of bubbles along the $x$ axis, which contribute to, respectively, the height of each potential drop and the number of occurrence of potential drop along the $x$ axis.
The dependence of $V_x^N$ on the number of bubbles and the sample geometry is demonstrated in Fig.~3 (a).
Comparing the configurations i and ii, which share the same sample shape, the slope of $V_x$ is twice larger for ii because the configuration ii contains twice as many bubbles, i.e., the sites where the potential drop occurs.
It is not necessarily that the same value of $N$ leads to the same magnitude of electric voltage;
the configuration iii provides twice larger $V_x^N$ than ii does under the same applied field because of the difference in the factor $1/L_y$.
Similar discussion is applied for $V_y^N$, see Fig.~3 (b).
Eq.~(\ref{vn}) indicates that one may control the dc electric voltages by adjusting the sample geometry and the number of bubbles.

\begin{figure}[b]
    \centerline{\includegraphics[width=80mm]{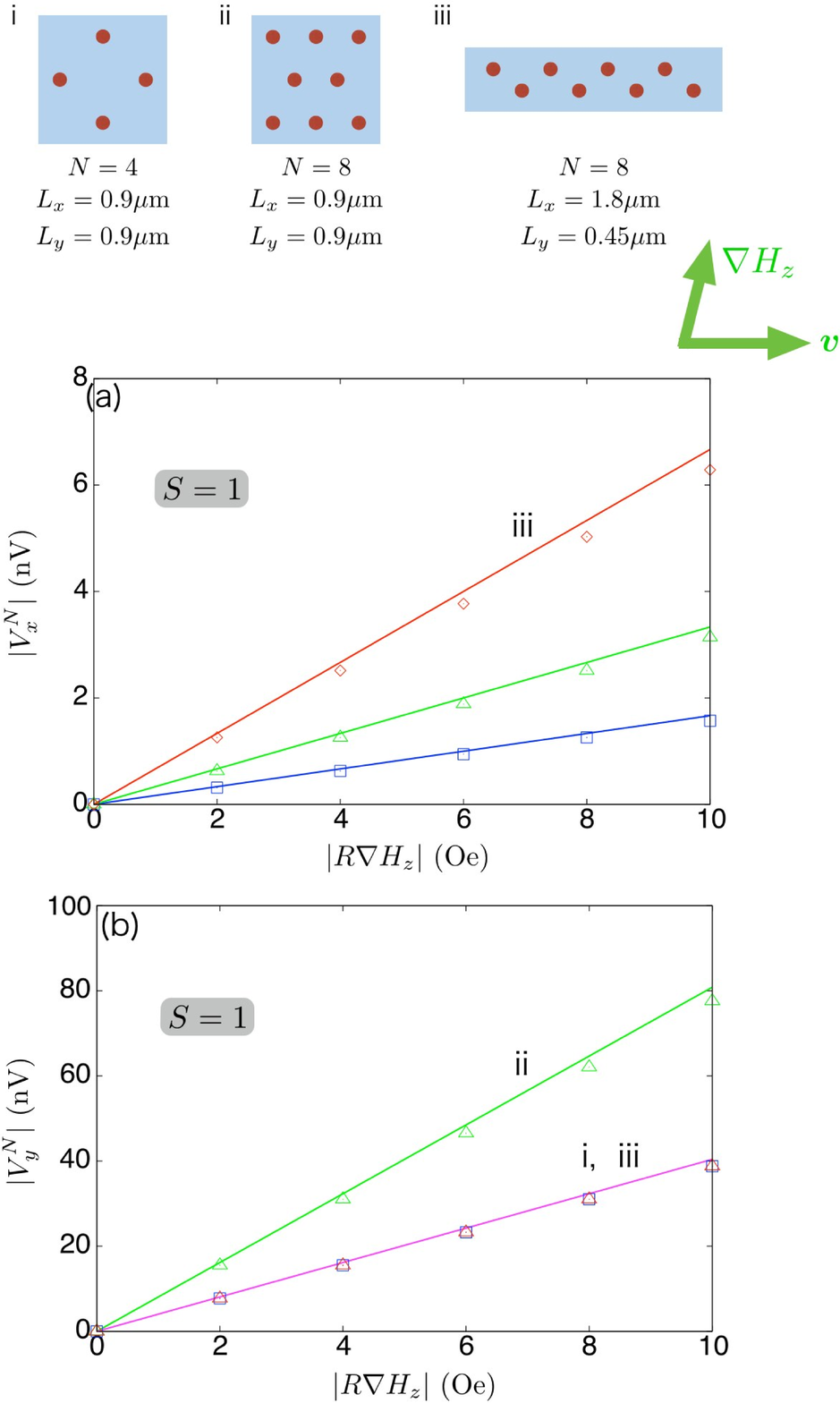}}
    \caption{  The field dependence of dc electric voltages induced by the steady motion of arrays of identical bubbles.
    Three different configurations are examined, which are indicated in the top panel.
    The dynamics of each bubble is described by Eqs.~(\ref{theta}), (\ref{psi}) and (\ref{steady}) with $S=1$ [see the caption of Fig.~2 for the other parameters assumed here].
    (a) $V_x^N$ obtained by Eq.~(\ref{vn}) and by numerically solving the Poisson equation~(\ref{poisson}) correspond to solid lines and open symbols, respectively, and (b) similarly for $V_y^N$.
    The electric voltages are proportional to the field gradient, and depends on the number of bubbles and the sample geometry.
    See also Discussion in the main text.
            }
    \label{fig2}
\end{figure}

Remarkably, only the non-adiabatic field $\bm{{\cal E}}^{\rm NA}$ contributes to $V_x^N$, while only the adiabatic one $\bm{{\cal E}}^{\rm A}$ to $V_y^N$.
Since in most magnetic materials $\beta$ is believed to be smaller than unity and thus $|\bm{{\cal E}}^{\rm NA}|\ll|\bm{{\cal E}}^{\rm A}|$, it is difficult to identify the contribution of the non-adiabatic field in the conventional systems.
In fact, there has been no experimental confirmation of its effects.
Now, by the measurement of $V_x^N$ one can detect $\bm{{\cal E}}^{\rm NA}$ free from the larger adiabatic contribution.
This can lead to direct and unambiguous measurement of the phenomenological parameter $\beta$;
as both $V_x$ and $V_y$ contain in their expressions $P$, which may be another uncertain material parameter, one can be free from $P$ by measuring the ratio of $V_x$ and $V_y$.

In the present study, we assumed the specific and simple profile of magnetization, Eqs.~(\ref{theta}), (\ref{psi}) and (\ref{steady}), to derive the analytic expressions for the electric voltages.
Although the investigation of the effects of the bubble shape distortion, local disorder, and etc. must be interesting, it will require systematic study based on numerical approaches to the magnetization dynamics, which is out of reach of this paper.
The assumption made in the present study is reasonable when the gradient of applied field is sufficiently moderate,\cite{bubble2} and our analytical model will work well in this field range. 

In conclusion, We have shown for the first time that a magnetic filed gradient can generate a SMF, i.e., spin currents and associated electric voltages, by driving the motion of magnetic bubbles.
Based on the steady-motion model, expressions for the dc electric voltages in longitudinal and perpendicular to the bubble motion are derived, which turned out to be controllable by tuning the sample geometry and the number of involved bubbles.
An important implication of our result is that the present system offers an experimental determination of the phenomenological parameter $\beta$ that describes non-adiabaticity in the electron spin dynamics.
This SMF can lead to a new route for basic study of the electron-magnetization interaction as well as a new concept in spintronic devices, exploiting the gradient of magnetic fields.
\\

\noindent {\bf Acknowledgements}
The authors would like to thank Cristian Cernov for valuable discussion and making the image for Fig.~1.
This research was supported by Research Fellowship for Young Scientists from Japan Society for the Promotion of Science, Grant-in-Aid for Scientific Research (No.~24740247 and No.~26247063) from MEXT, Japan, and Alexander von Humboldt Foundation.


\newpage
\noindent {\bf Appendices}

\noindent
{\bf Steady-motion of bubble.}
In order to get to the equation of motion for the bubble, consider the increment of the stored magnetic energy $U=\int w\ dV$, with the magnetic energy density $w$, due to the variations $\delta \theta$ and $\delta\psi$:
\begin{eqnarray}
\delta U&=& \iiint  \left(\frac{\delta w}{\delta\theta}\delta\theta+\frac{\delta w}{\delta\psi}\delta\psi\right) dxdydz \nonumber\\
 &=&-\frac{2\mu_0 M_{\rm S}|{\bm v}|\pi Rh}{\gamma}\left[\frac{\alpha}{\Delta}\left( 1 + \frac{S^2\Delta^2}{R^2}\right)dX+\frac{2 QS}{R}dY\right],\nonumber\\
\label{bubble_w}\end{eqnarray}
where $h$ is the film thickness and $M_{\rm S}$ is the saturation magnetization.
$\delta w/\delta\theta$ and $\delta w/\delta\psi$ have been expressed in terms of $\partial\theta/\partial t$ and $\partial\psi/\partial t$ by using Landau-Lifshitz-Gilbert equation of motion without spin-transfer-torque effect, and the time derivatives have been further expressed in terms of $r$ and $\chi$ assuming Eqs.~(\ref{theta}), (\ref{psi}) and (\ref{steady}).
The change $\delta U$ in the internal energy is supposed to be balanced by the external pressure on the bubble due to $\nabla H_z$:
\begin{equation}
Q\left(2\mu_0 M_S \nabla H_z\right)\pi R^2 h = -\frac{dU}{d{\bm X}},
\end{equation}
where ${\bm X}$ denotes the position of the center of the bubble.
Rearranging the above equation, Eq.~(\ref{eom}) is obtained.
The net mobility of the bubble is found by solving Eq.~(\ref{eom}) for $|{\bm v}|$ in terms of $|R\nabla H_z|$ as 
\begin{equation}
|{\bm v}| = \frac{R\gamma}{2|S|} \left[ 1+\alpha^2\left(\frac{1+\left(S\Delta/R\right)^2}{2S\Delta/R}\right)^2 \right]^{-1/2} |R\nabla H_z|.
\end{equation}
The angle $\rho$ of deflection of the bubble away from the field gradient may be defined as
\begin{equation}
\rho = \tan^{-1}\frac{\partial H_z/\partial y}{\partial H_z/\partial x} 
= \cot^{-1}\frac{Q\alpha\left\{ 1+ (S\Delta/R)^2\right\}}{2S\Delta/R}.
\end{equation}
With $S=1$, $Q=1$, $\alpha=0.02$, $\Delta=2$ nm and $R=50$ nm, corresponding to the calculation in Fig.~2, one obtains $\rho\simeq75.9^\circ$.
\\

\noindent
{\bf Derivation of spin electric fields.}
Under the Hamiltonian (\ref{h}), the Heisenberg equation of motion for the electron is given by
\begin{equation}
\bm{{\cal F}}=[[{\bm r},{\cal H}],{\cal H}]/(i\hbar)^2=-J_{\rm ex}{\bm \sigma}\cdot\nabla{\bm m},
\end{equation}
with ${\bm r}$ and $\bm{{\cal F}}$ denoting the operators for the electron's position and the force acting on the electron, respectively.
The actual motion of the electron is obtained by determining the expectation value $\langle{\bm \sigma}\rangle_{\uparrow\downarrow}$ of the electron spin with majority ($\uparrow$) and minority ($\downarrow$) states.
Notice that ${\bm m}\cdot\nabla{\bm m}=0$ and thus the component of $\langle{\bm \sigma}\rangle_{\uparrow\downarrow}$ that is (anti-)parallel to ${\bm m}$ does not contribute to the force.
Let us decompose the electron spin as $\langle{\bm \sigma}\rangle_{\uparrow\downarrow}=\mp{\bm m}+\delta{\bm m}_{\uparrow\downarrow}$, where the upper (lower) sign corresponds to the majority (minority) spin and $\delta{\bm m}_{\uparrow\downarrow}$ represents a slight deviation from $\mp{\bm m}$.
The expectation value of the force is written as $\langle\bm{{\cal F}}\rangle_{\uparrow\downarrow} = - J_{\rm ex}\delta{\bm m}_{\uparrow\downarrow}\cdot\nabla{\bm m}$;
what causes non-zero force due to the exchange coupling is a misalignment between the electron spin and the magnetization.
Assume that the electron spin dynamics is described by
\begin{equation}
\frac{\partial\langle{\bm \sigma}\rangle_{\uparrow\downarrow} } { \partial t }
= - \frac{ 2J_{\rm ex} }{ \hbar } \langle{\bm \sigma}\rangle_{\uparrow\downarrow} \times {\bm m}
- \frac{ \delta{\bm m}_{\uparrow\downarrow} } { \tau_{\rm sf} }.
\label{eom-spin}\end{equation}
The first term on the right-hand side represents the Larmor precession about the magnetization, and the damping motion toward the magnetization is phenomenologically introduced by the second term, which describes the non-adiabaticity in the electron spin dynamics, with $\tau_{\rm sf}$ the relaxation time for the electron spin flip.
By substituting the above expression for $\langle{\bm \sigma}\rangle_{\uparrow\downarrow}$ into Eq.~(\ref{eom-spin}), $\delta{\bm m}_{\uparrow\downarrow}$ is expressed in terms of ${\bm m}$, by which one obtains $\langle\bm{{\cal F}}\rangle_{\uparrow\downarrow}=\pm(-e\bm{{\cal E}})$ with $\bm{{\cal E}}$ given by Eq.~(\ref{es}).

We have considered an open circuit condition.
More generally Eq.~(\ref{eom-spin}) should include the divergence of the spin current carried by conduction electrons, leading to appearance of the spin magnetic field.
While we have followed Ref.~[\cite{prb}] here, essentially the same result was obtained by a different approach where Onsager's reciprocal relation is taken into account.\cite{tserkov,duine}
\\

\noindent
{\bf Numerical approach to the Poisson equation.}
In the numerical calculations, the sample is divided into number of meshes, and a single magnetization vector ${\bm m}_i$ is assigned to each mesh, where $i$ is the index of the meshes.
The time evolution of $\{ {\bm m}_i(t) \}$ $(i=1,2,...,N_{\rm m})$, where $N_{\rm m}$ is the number of the meshes, is 
given based on Eqs.~(\ref{theta}), (\ref{psi}), and (\ref{steady}).
As the spin electric field $\{\bm{{\cal E}}_i\}$ is updated at each mesh by Eq.~(\ref{es}), the induced electric potential distribution $\{\phi_i(t)\}$ satisfies the Poisson equation [see Eq.~(\ref{poisson})]
\begin{equation}
\nabla^2 \phi_i(t) = P \nabla\cdot\bm{{\cal E}}_i(t).
\end{equation}
In two-dimensional discrete systems, the above equation is equivalent to 
\begin{equation}
\phi_{i,j} = \frac{1}{4}\left(\nabla\cdot\bm{{\cal E}}_{i,j} \Delta x \Delta y + \phi_{i+1,j} + \phi_{i-1,j} + \phi_{i,j+1} + \phi_{i,j-1} \right),
\end{equation}
where ($i, j$) are the indices for the meshes in two dimension, and $\Delta x\Delta y$ is the area of the mesh.
The potential distribution is obtained by solving the above equation self-consistently with Neumann boundary condition, where the spatial derivative of the electric potential is zero at the sample edge.

\end{document}